# Bare and Polymer Coated Iron Oxide Superparamagnetic Nanoparticles for Effective Removal of U (VI) from Acidic and Neutral Aqueous Medium

*Shan Zhu[1,2,⊥], Yangchun Leng[2,3,⊥], Minhao Yan[*,2], Xianguo Tuo[*,3], Jianbo Yang[1], László Almásy[2,4], Qiang Tian[2], Guangai Sun[5], Lin Zou[5], Qintang Li[2], Jérémie Courtois[2] and Hong Zhang[2]*

1 The College of Nuclear Technology and Automation Engineering, Chengdu University of Technology, Chengdu 61009, China

2 State Key Laboratory Cultivation Base for Nonmetal Composites and Functional Materials, Southwest University of Science and Technology, Mianyang 621010, China

3 Sichuan University of Science and Engineering, Zigong 643000, China.

4 Institute for Solid State Physics and Optics, Wigner Research Centre for Physics, POB 49, Budapest, 1525, Hungary

5 Key Laboratory of Neutron Physics and Institute of Nuclear Physics and Chemistry, China Academy of Engineering Physics (CAEP), Mianyang 621999, China

⊥ S. Z and Y. L contributed equally to this work.

* to whom correspondence should be addressed:
- Prof. Yan Minhao
  State Key Laboratory Cultivation Base for Nonmetal Composites and Functional Materials
  Southwest University of Science and Technology (SWUST),
  Mianyang
  621010
  P. R. China
  Tel : 008613778092806
  Email : yanminhao@swust.edu.cn

- Prof. Tuo Xianguo
  Sichuan University of Science and Engineering
  Zigong 643000
  China
  Email: tuoxianguo@suse.edu.cn

**Abstract:**

Superparamagnetic $\gamma$-$Fe_2O_3$ nanoparticles (5 nm diameter) were synthesized in water. The bare particles exhibit good colloidal stability at ~ pH 2 because of the strong electrostatic repulsion with a surface charge of +25 mV. The polyacrylic acid (PAA)-coated particles exhibit remarkable colloidal stability at ~ pH 7 with abundant free carboxyl groups as reactive sites for subsequent functionalization. In this work, we used zeta potential analysis, transmission electron microscopy, small angle X-ray scattering, and Inductively coupled plasma mass spectrometry to investigate the adsorption behavior of U (VI) on bare and coated colloidal superparamagnetic nanoparticles at pH 2 and pH 7. At pH 2, uranyl ion ($UO_2^{2+}$) absorbed on the surface of the bare particles with decreasing particle surface charge. This induced particle agglomeration. At pH 7, uranyl ion ($UO_2^{2+}$) hydrolyzed and formed plate-like particles of uranium hydroxide that were ~ 50 nm in diameter. The PAA-coated iron oxide nanoparticles absorbed on the surface of these U (VI) hydroxide plates to form large aggregates that precipitate to the bottom of the dispersion. At both pH 2 and pH 7, the resulting U (VI)/nanoparticle complex can be easily collected and extracted from the aqueous environment via an external magnetic field. The results show that both bare and polymer-coated superparamagnetic $\gamma$-$Fe_2O_3$ nanoparticles are potential absorbents for removing U (VI) from water.

**Introduction**

Uranium sequestration is an important goal due to prior weapons production, uranium ore mining process, spent fuel reprocessing, and the increasing interest in nuclear power. Uranium is radioactive and toxic and mainly exists as U (IV) and U (VI); U (IV) is a less soluble form and is only found in relatively reducing environments. U (VI) is the most common oxidation state, and it exists as the uranyl ion ($UO_2^{2+}$) with different aqueous complexes of uranyl including hydroxide and carbonate. Tools to remove uranium (VI) from aqueous solutions include chemical precipitation[1], membrane dialysis[2], solvent extraction[3], flotation[4], and adsorption [5-8].

Extraction[9, 10] and adsorption[11] are the most valid ways to eliminate U (VI) from aqueous solutions. Adsorption is simple and has little secondary pollution, and numerous absorbents such as carbon[12], bentonite[13] and polymers[14] have been developed. To more easily extract adsorbed U (VI) from water, a variety of magnetic adsorbents such as magnetite[15], $Fe_3O_4$@$TiO_2$[11], or novel magnetite nanoparticles containing calix arenes[16] have recently been used as a host absorbent to sequester the U (VI). The strategy is usually to functionalize these magnetic adsorbents with a specific chelator to $UO_2^{2+}$ and then extract the adsorbed U (VI) with a magnetic adsorbent and an external magnetic field.

This strategy works well at lower pH (pH < 5.5), but the ligand groups are usually deprotonated at increasing pH. This increases the effective chelating sites. However, $UO_2^{2+}$ hydroxide can form at ~ pH 7. This is not conducive to chelation, and this limits practical applications of extraction/adsorption in neutral media[17].

Here, we synthesized highly stable 5 nm colloidal $\gamma$-Fe$_2$O$_3$ superparamagnetic nanoparticles (NPs). We then coated them with polyacrylic acid (PAA), which offers remarkable colloidal stability (years) as well as reactive carboxyl groups for subsequent functionalization. The bare and coated particles could absorb U (VI) at pH 2 and pH 7, respectively. Small-angle X-rays scattering (SAXS) was used to investigate the adsorption of U (VI) in these colloidal system via a combination of zeta potential analysis, transmission electron microscopy, and ICP-MS. The adsorption mechanism at pH 2 and pH 7 is also discussed. The results show that bare and PAA-coated particles can absorb U (VI) with good performance (the distribution coefficient $K_d >10^3$ ml/g) in both acidic and neutral pH environment. Moreover, the absorbed complexes can be extracted from aqueous media via an external magnetic field.

## 2. Materials and Methods

### 2.1 Materials

FeCl$_2$·4H$_2$O, FeCl$_3$·6H$_2$O, Fe(NO$_3$)$_3$·9H$_2$O, ammonia (20 wt.% in water), nitric acid (52 wt.% in water), acetone, diethyl ether, and poly(acrylic acid) with $M_W$ = 15000 g. mol$^{-1}$ were obtained from Sigma Aldrich and used as received. MilliQ quality water (Millipore®) was used throughout. Standard solutions of U (VI) (GBW(E)080173; uranyl nitrate in pH 0.2 with concentration of 100 μg/ml) were obtained from Beijing Research Institute of Chemical Engineering and Metallurgy.

### 2.2 Synthesis of nanoparticles

Maghemite ($\gamma$-Fe$_2$O$_3$) nanocrystals were synthesized via the Massart method[18]. Briefly, iron (II) and iron (III) salts were co-precipitated in alkaline aqueous media at room temperature. The resulting magnetite ($Fe_3O_4$) nanocrystals were then transferred into an acidic aqueous medium using nitric acid. This was then oxidized into maghemite via addition of $Fe_3(NO_3)_3$ with boiling solvent. This oxidized the magnetite $Fe_3O_4$ into stable $\gamma$-$Fe_2O_3$ maghemite nanoparticles. The crystalline nature of the maghemite nanoparticles was characterized with electron diffraction (Figure 1). The pattern is characteristic of maghemite ($\gamma$-$Fe_2O_3$).

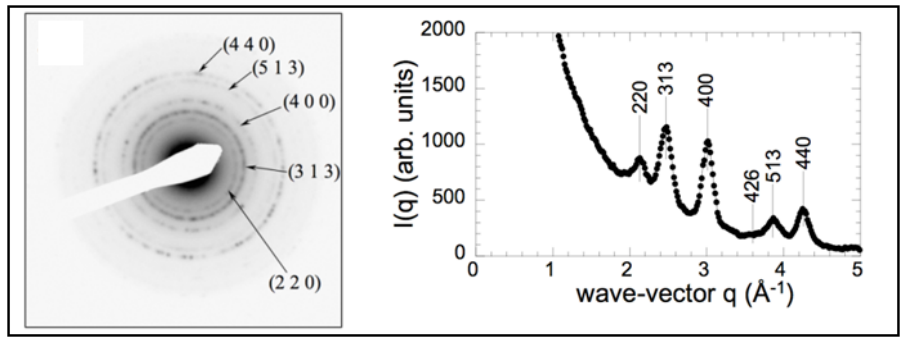

*Figure 1: Microdiffraction spectra of maghemite nanocrystals.*

Next, we performed vibrating sample magnetometry (VSM) measurements of a liquid dispersion of $\gamma$-$Fe_2O_3$. The VSM curve shown in Figure 2 evidenced superparamagnetic behavior of the obtained NPs at T = 25°C with a saturation magnetization of 46.2 emu. $g^{-1}$. The synthesis gave maghemite nanoparticles with a broad distribution of sizes. More than 250 nanoparticles with clear edge from different areas in TEM sample grid were selected and counted [19]. The data were then fitted using a log-normal distribution function (equation (1)) [20, 21] with median diameter $D_0^{TEM}$ = 10 ± 0.5 nm and polydispersity $s^{TEM}$ = 0.4 ± 0.05:

$$p(D) = \frac{1}{\sqrt{2\pi}(s^{TEM})D} \exp\left(-\frac{\ln^2(D/D_0^{TEM})}{2(s^{TEM})^2}\right) \qquad (1)$$

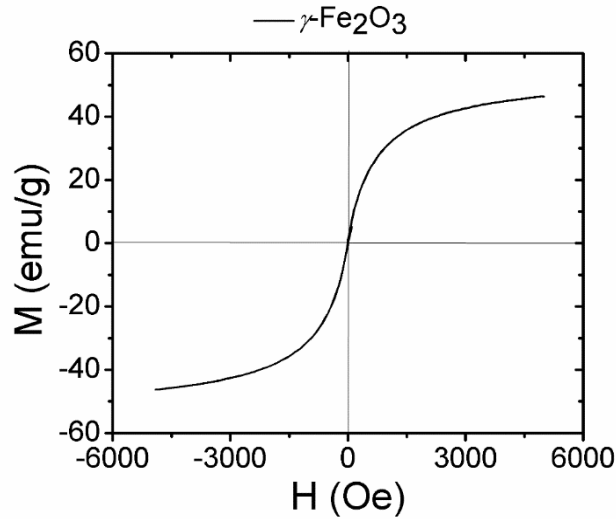

*Figure 2: Magnetic field dependence of the macroscopic magnetization M(H) for γ-Fe$_2$O$_3$ NP dispersions.*

Size sorting was used to reduce the polydispersity of the γ-Fe$_2$O$_3$ NPs using successive liquid-liquid phase separations induced by addition of nitric acid[22]. Adding large amounts of nitric acid decreased the pH from 2 to 0.5 and increased the ionic strength. The phase separation is liquid-gas, and the more concentrated phase was separated via magnetic sedimentation. The concentrated phase was called "C", and the diluted phase is "S" (for supernatant). Prior studies [18, 23] showed that the concentrated phase contained the largest particles, and the diluted phase contains the smallest ones.

The flocculated particles in "C" were then redispersed by adding water to increase the pH from 0.5 to 1.8. This also decreases the ionic strength via dilution. The "S" phase contains dispersed particles and counterions (especially nitrate, NO$_3^-$) that must be removed. In this way, the γ-Fe$_2$O$_3$ NPs are sorted in two batches: "C" and "S" containing larger and smaller particles, respectively. The process of sorting using phase separation can be repeated on "C" leading to C1C and C1S ("S1C" and "S1S" can be prepared similarly). Ultimately, γ-Fe$_2$O$_3$ nanocrystals with nominal diameters of 5 nm suspended in acidic aqueous media (pH 1.8) were obtained with polydispersity of 0.18.

Next, the bare NPs were coated in an acidic environment with 15000 g·mol$^{-1}$ PAA using the precipitation-redispersion process[24]. The drop-by-drop addition of a solution of PAA at pH = 1.8 led to a dispersion of bare NPs at pH = 1.8. This resulted in precipitation of the nanocrystals with polymer chains adsorbed to the surface. Here the mass ratio between added PAA and bare NPs was 10:1 in order to achieve coating saturation.

Single particles were then recovered by redispersion at pH = 10 with NH$_4$OH. The PAA-coated NPs were purified via dialysis in deionized water to remove the excess un-coated PAA and other impurities. The coating process is illustrated in Scheme 1. The resulting PAA$_{15K}$-coated NPs (denoted NP-PAA$_{15K}$) were stored in neutral aqueous media (in H$_2$O at pH 7). The stability and resilience of the poly(acrylic acid) coating have been investigated previously[25]. These results confirm the presence of a highly resilient PAA adlayer on the $\gamma$-Fe$_2$O$_3$ nanoparticles—this is crucial for many applications. Moreover the electrostatic density *e.g.* the average density of carboxylic acid groups around NPs was calculated to be $n^{coo-}$ = 28 nm$^{-2}$ with a percentage of chargeable coo$^-$ of 60% by using the method reported previously[25]. These results show that the present coating method ensures a uniform and dense coating of the particles, with an elevated and constant density of chains and charges.

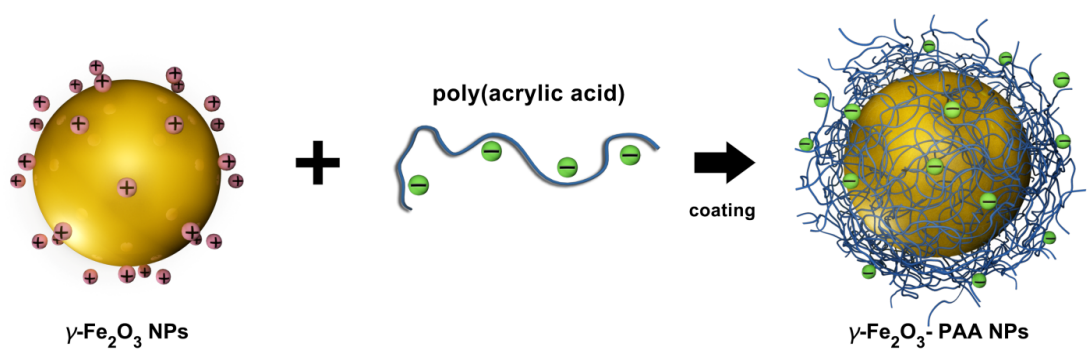

***Scheme 1.*** *Representation of the coating process using positively charged bare NPs and negatively charged PAA oligomers.*

**2.3 Adsorption**

Centrifuge tubes were cleaned in an ultrasonic bath using warm (40°C) water for 30 min. They were then washed twice with normal tap water, soaked in double distilled water for 2 h, and dried at 40°C. The NP dispersion (2000 μL) was placed into centrifuge tubes, and aliquots (200 μL, 400 μL, 600 μL, 800 μL, and 1000 μL) of U (VI) solution were added. Before the NPs were mixed with U (VI) solution, 1 wt.% $NH_3·H_2O$ was added to the $UO_2^{2+}$ solution to increase the pH from 0.25 to 2 and 7—these values correspond to the pH of the bare and coated NP dispersions.

The tubes were then shaken by constant temperature shaker at 25 °C with speed of 250 rpm for 48 hours in order to reach sorption equilibrium. Afterwards, the tube containing bare NPs was centrifuged for 2 h at 10000 rpm. After centrifugation, the precipitate at the bottom of the centrifugation tube was collected with an external magnetic field, and 300 μL of the supernatant was transferred to a quartz cuvette for subsequent concentration characterization using Inductively coupled plasma mass spectrometry (ICP-MS). Aggregates of the NP-$PAA_{15K}$ precipitate were collected with an external magnetic field. We then decreased the pH of the residual dispersion from 7 to 0.5 and characterized the concentration of U (VI) after adsorption with ICP-MS (Agilent 7700x). The sample was maintained at 25°C throughout this process. The adsorption experiments were conducted under $N_2$ to exclude the complexation of uranyl by dissolved carbonate. A control was used to estimate the adsorption of U (VI) (about 10%) by the centrifuge tube.

## 2.4 Characterization

The zeta potential was monitored on a multi-angle particle size and zeta potential analyzer (Brookhaven NanoBrook Omni). Transmission electron microscopy (TEM) used a Zeiss Libra200FE at the Analysis and Characterization Center of Southwest University of Science and Technology. The U (VI) concentration was measured with ICP-MS (Agilent 7700x).

The X-ray scattering measurements were performed by a SAXSpace small angle X-ray scattering instrument (Anton Paar, Austria, Cu-K$\alpha$, $\lambda$ = 0.154 nm) equipped with a Kratky block-collimation system and an image plate as the detector. The X-ray generator was operated at 40 kV and 50 mA. A temperature control unit Anton-Paar TCS 150 connected with the SAXSpace maintaned the temperature at 25°C. Samples were transferred to thin-wall quartz capillaries with an inner diameter of 1 mm. For colloidal NP samples, 1-hour exposure times were used to obtain good signal-to-noise ratios. The scattering curve of pure $H_2O$ in the same capillary served as the background (also with a 1-hour exposure time). All the data were corrected for transmission and background scattering of the capillary and $H_2O$.

## 3. Results and Discussion

### 3.1 Adsorption in acidic environment (pH 2)

The U (VI) adsorption of bare NPs was first investigated at pH 2. In this case, the U (VI) stays as $UO_2^{2+}$ to complex with the surface hydroxyl groups on the bare NPs. Thus, the adsorption mechanism is surface complexation between the $OH^-$ group and the $UO_2^{2+}$ ions.

TEM images (Figures 3) of the bare NPs before and after adsorption of $UO_2^{2+}$ show the following features: 1) The bare NPs remain stable and monodisperse before adsorption as expected; 2) The particles agglomerate after adsorption. At pH 2, the cations are adsorbed on the bare NPs despite the electrostatic repulsion caused by a positive surface charge. Before adsorption, the positive surface charges (+ 25 mV) ensure colloidal stability of bare NPs [26]. After adsorption, the surface charge of the NPs decreased to + 7 mV, indicating that the number of cations around the surface of NPs became much less. Under these acidic conditions, the $UO_2^{2+}$ ions complex with the surface hydroxyl group (denoted as =FeOH) at the surface of NPs:

$$2(=FeOH) + UO_2^{2+} \leftrightarrow UO_2(OFe=)_2 + 2H^+$$

Therefore, adsorption in this case reflects the complex solution chemistry of the uranyl ion. The adsorption of $UO_2^{2+}$ on the NP surface can destroy the electrostatic balance between the surface hydroxyl group $OH^-$ and the $H^+$ close to the particle surface, thus decrease the number of cations near the NPs surface. This induces particle agglomeration.

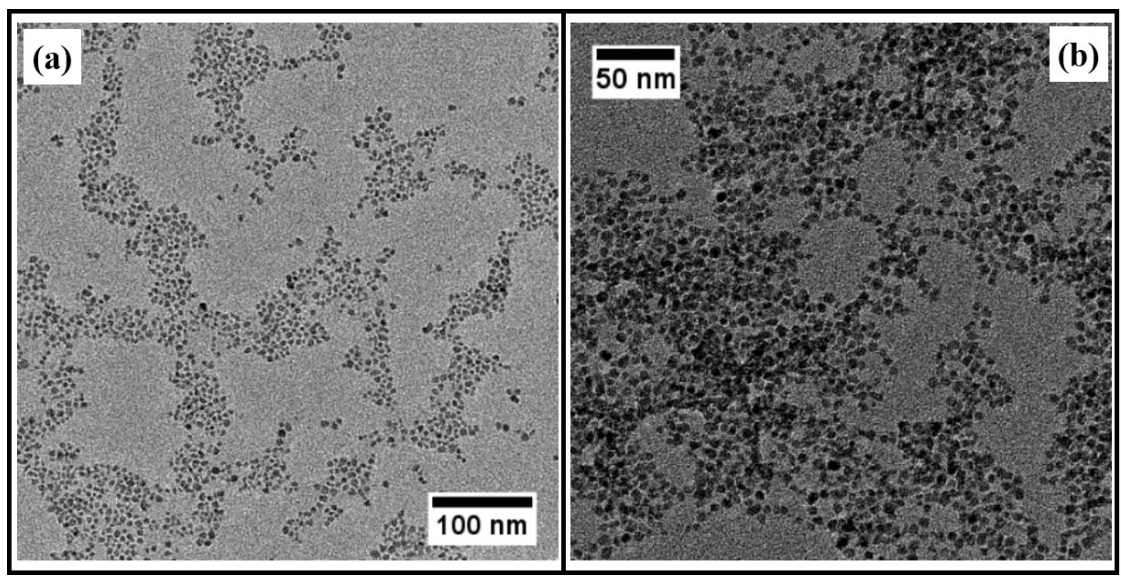

*Figures 3. TEM images of bare NPs before (a) and after (b) adsorption of $UO_2^{2+}$.*

Next, SAXS was used to get more detailed information about the colloid. The SAXS data before adsorption does not show agglomeration and is easily fitted via the spherical model with a lognormal size distribution (Figure 4). The scattering intensity can be written as:,

$$I(q) = \Delta\rho^2 \int_0^\infty (\frac{4\pi}{3}R^3)^2 N(R)P(q, R)dR \qquad (2)$$

$$N(R) = \frac{N_0}{\sigma R\sqrt{2\pi}} \exp\left(-\frac{(\ln R - \ln R_{med})^2}{2\sigma^2}\right) \qquad (3)$$

$$P(qR) = \left[\frac{3(\sin(qR) - qR\cos(qR))}{(qR)^3}\right]^2 \qquad (4)$$

Here, $\Delta\rho$ is the difference in electron scattering length density between the aqueous solution and $\gamma$-$Fe_2O_3$ particles, $R$ is the particle radius, $N_0$ is the total number of particles per unit volume, $R_{med}$ is the median radius, $\sigma$ is the width of the size distribution, $\sigma$ is the polydispersity parameter, and $P(qR)$ is the form factor of spheres. The fitted median diameter is 4.98 nm with a polydispersity index of 0.3. The intensity at low $q$ region increases after adsorption indicating NP agglomeration. In this case, the data cannot be fitted via the individual particle model.

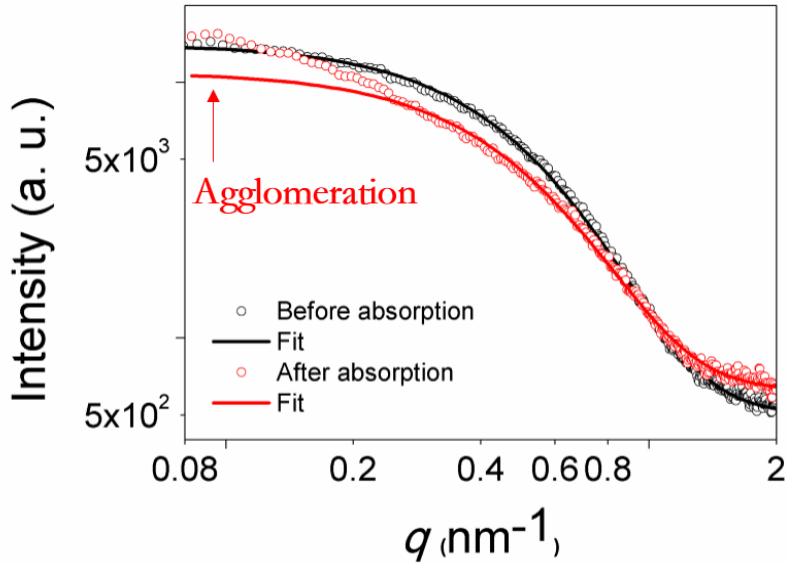

Figure 4. SAXS data of bare NPs before and after adsorption of $UO_2^{2+}$. The continuous lines are the best fits to the log-normal size distribution of spherical

*particles.*

## 3.2 Adsorption in neutral environment (pH 7)

Uranyl species are oligomeric above pH 5 and at moderate initial $UO_2^{2+}$ concentrations. They are mainly $(UO_2)_3(OH)_5^+$ species due to hydrolysis.[27, 28]

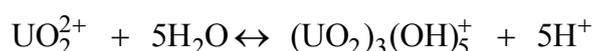

$$UO_2^{2+} + 5H_2O \leftrightarrow (UO_2)_3(OH)_5^+ + 5H^+$$

At neutral pH ~ 7, the $UO_2^{2+}$ hydrolyze and form plate-like hydroxide aggregates (Figures 5). The plate-like hydroxide has a diameter of 30 ~ 50 nm via TEM. The surface charges of these plate-like hydroxides is + 5 mV. SAXS measured on dispersion containing these hydroxides showed a power law in the $q$ range of 0.06 to 0.2 nm$^{-1}$. Fitting showed that this power law is $q^{-2}$ (Figure 5 solid line). This power law indicates formation of plate-like structure[29]. Due to the limited $q$ range and polydispersity of the $(UO_2)_3(OH)_5^+$, it is diffcult to derive the dimensions of $(UO_2)_3(OH)_5^+$ hydroxide by a theoretical model. At neutral pH, these U (VI) oligomers can interact with the anionic PAA-coated NPs via electrostatic complexation.

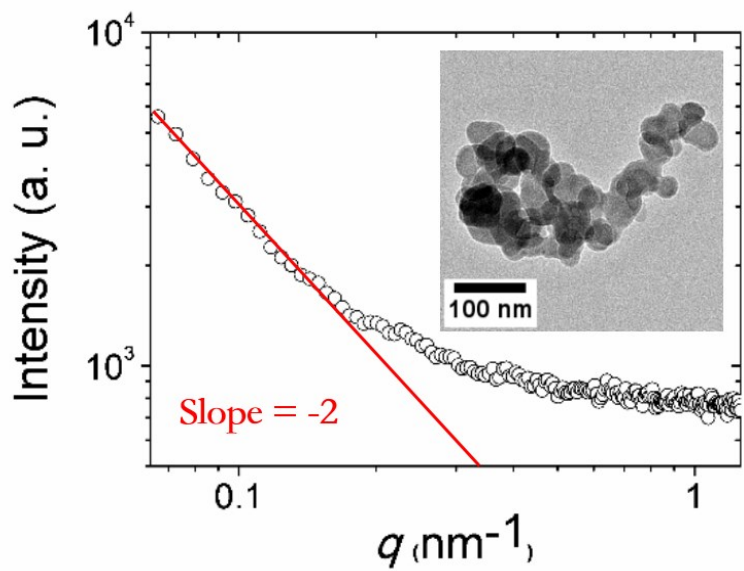

*Figures 5. Experimental SAXS pattern of the U (VI) hydroxide aqueous dispersion. Solid line shows the $q^{-2}$ power law characteristic of plate-like structures. Inset: TEM image of U (VI) hydroxide shows plate-like structures consistent with the SAXS data.*

These hydroxides cannot easily capture the chelator in neutral pH.[30] Here the poly(acrylic acid)-coated NPs were employed to absorb/capture these hydroxide plates. The TEM images clearly show that the polymer-coated NPs are individually dispersed before adsorption at neutral pH (Figures 6a). The surface charge of these NP-PAA$_{15K}$ is -39 mv, which ensures excellent colloidal stability. After adsorption, the surface charge of species in the dispersion decreased to -5 mv. The TEM data clearly show that the polymer-coated NPs absorbed these plate-like hydroxides. This caused them to quickly sediment to the bottom of the sample holder. They can then be easily collected via an external magnetic field (Figures 6b).

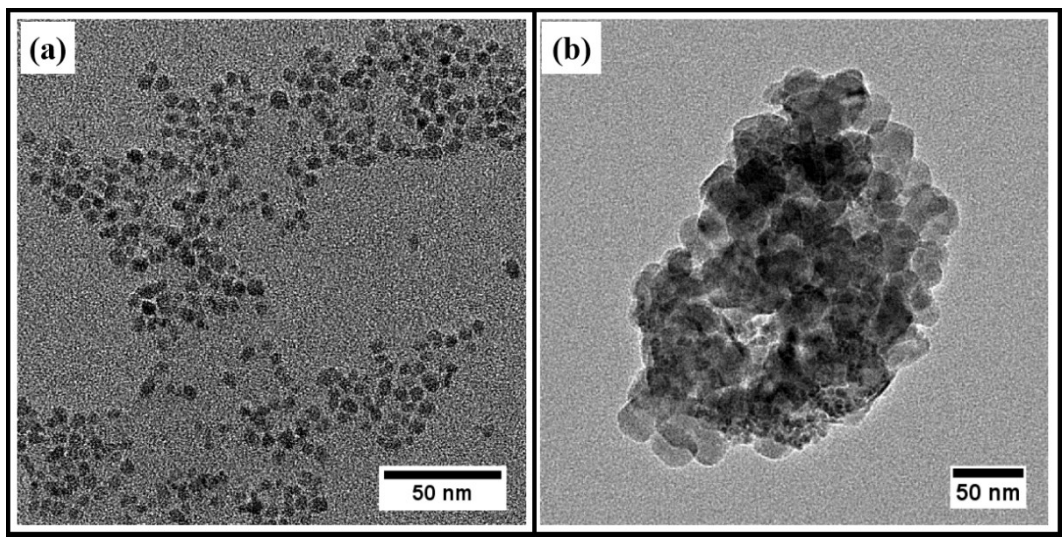

*Figures 6. TEM images of polymer coated NPs before (a) and after (b) adsorption of U (VI) hydroxide.*

SAXS revealed the adsorption behavior of these coated NPs. The SAXS data before adsorption show individual polydisperse particles that can be easily fitted via the spherical model with a log-normal size distribution (Figure 7). The fitted median diameter is 5.1 nm with a polydispersity of 0.3. The SAXS technique cannot discriminate the organic corona and inorganic core because the electron density contrast between PAA and the surrounding solvent is very small.[19] The intensity was much lower after adsorption. This means that the concentration of the iron oxide nanoparticles in the irradiated sample volume decreased. This decrease originated from the large amount of NPs-PAA$_{15k}$ absorbed on the surface of the plate-like U (VI) hydroxide, which precipitated and sedimented at the bottom of the sample holder.

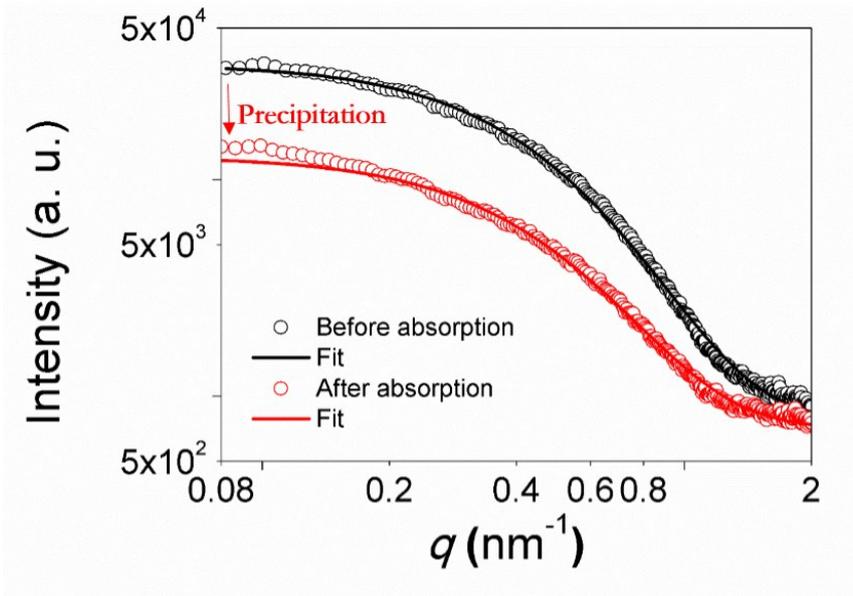

*Figure 7. SAXS data of NPs-PAA$_{15K}$ before and after adsorption of U (VI) hydroxide. The continuous lines are the best fits to the log-normal size distribution of spherical particles.*

**3.3 Adsorption performance**

The adsorption capacity of these bare and coated particles was then quantitatively investigated by calculating their distribution coefficient of U (VI) from solution. The distribution coefficient, $K_d$, is a measure of the distribution of the U (VI) between two phases and is exemplified here for particles and water. Higher $K_d$ values suggest that the solid is more effective at removing U (VI) from solution (**Equation (6)**):

$$K_d = \frac{C_0 - C_t}{C_t} \times \frac{V}{m} \qquad (6)$$

Here, $K_d$ is the sorption distribution ratio in mL g$^{-1}$, $C_0$ and $C_t$ are the concentrations of U (VI) in the solution before and after contact with the solid phase in mol dm$^{-3}$, $V$ is the volume of the aqueous phase in mL, and $m$ the mass of the solid phase in grams. The behavior of U (VI) collection by bare and coated NPs as a function of U (VI) concentration is shown in Figure 8.

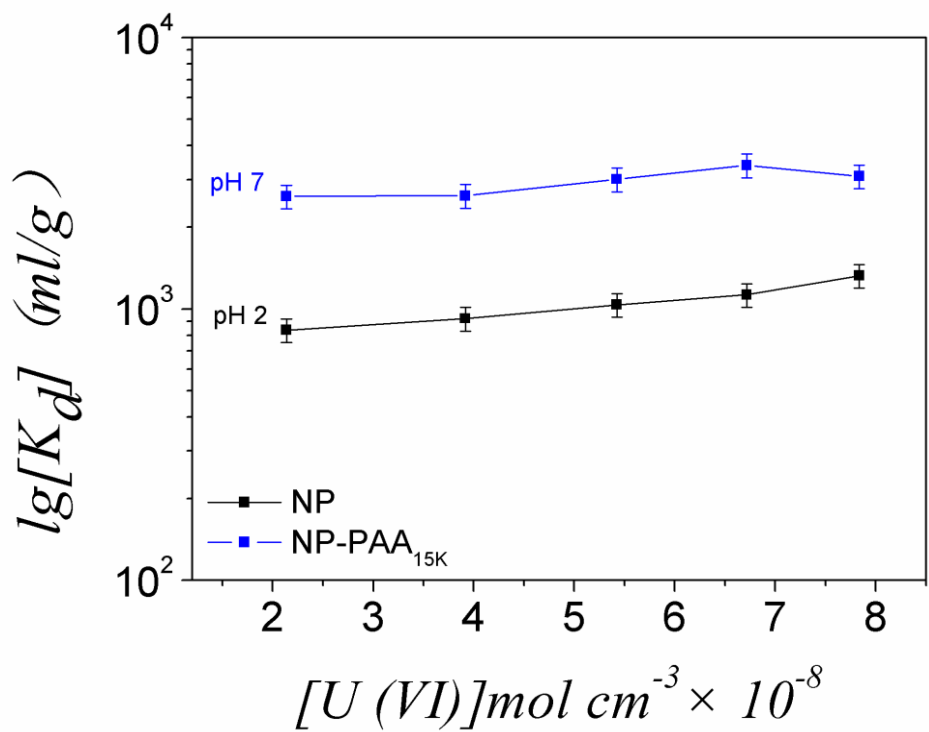

*Figure 8. Sorption distribution coefficient ($K_d$) versus U (VI) concentration obtained for bare and coated NPs at pH 2 and pH 7, respectively.*

The $K_d$ increased with U (VI) concentration for both bare and polymer-coated NPs. The $K_d$ of both bare and coated NPs are relevant (~ $10^3$ ml/g) indicating effective adsorption of U (VI) in both acidic and neutral pH. The polymer-coated NPs have better adsorption capability for these U (VI) hydroxides, which are usually incapable of being captured by the chelator. Moreover, these absorbed/captured U (VI) can be easily extracted with an external magnetic field.

We also studied the influence of size sorting on the adsorption performance. In pH 2 and at the highest U (VI) concentration ($7.84 \times 10^{-8}$ mol cm$^{-3}$) in the previous concentration range, the $K_d$ from original bare NPs was compared with the one from size sorted bare NPs. As shown in Figure 9, size sorted bare NPs present higher adsorption ability than the original one. It is reasonable because the decrease of both median diameter $D_0^{TEM}$ and polydispersity $s^{TEM}$ of NPs leads to higher surface area after size sorting, which brings higher adsorption capacity. This result is in agreement with the previous studies on the adsorption of U(VI) on soil minerals: extent of U(VI) sorption depends more on reactive surface area than on the surface properties of sorbents.[31]

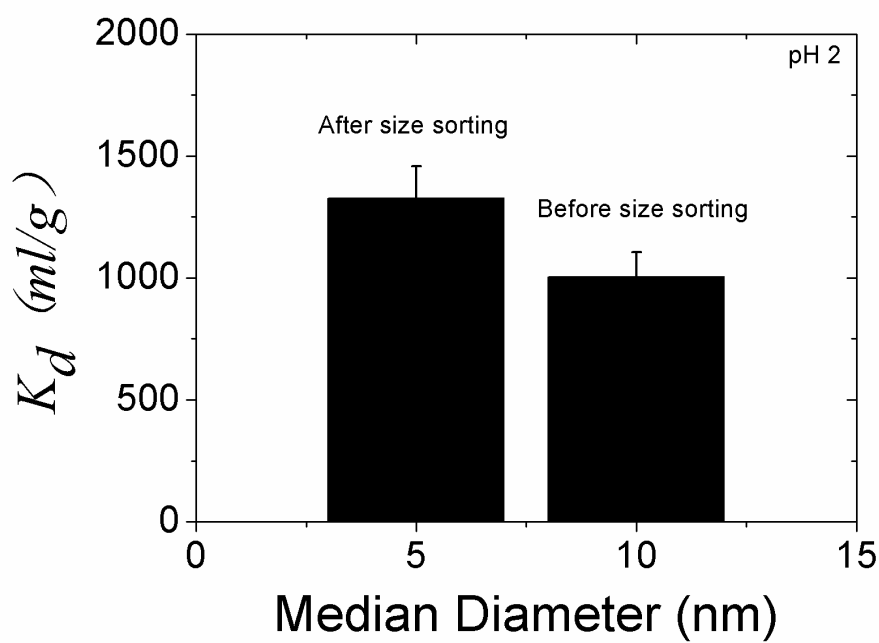

*Figure 9. Sorption distribution coefficient ($K_d$) from bare NPs before and after size sorting in pH 2 and at U (VI) concentration of $7.84 \times 10^{-8}$ mol cm$^{-3}$.*

Because of the existence of various metal ions in natural water, we then investigate the adsorption ability of bare and coated NPs on different metal ions such as $Sr^{2+}$, $Ba^{2+}$, $Ca^{2+}$, $VO_3^-$, $Na^+$, $Mg^{2+}$, and $Zn^{2+}$ to evaluate the selectivity capacity of these NPs. For each of these metal ions, their concentration was maintained at $7.84 \times 10^{-8}$ mol cm$^{-3}$ which is the same as U (VI) highest concentration in the previous experiment. Then the sorption distribution coefficient ($K_d$) of bare and coated NPs on each of these metal ions was measured in pH 2 and pH 7, respectively. Figure 10 displays the result, and Figure 10 (a) shows that there is almost no difference between the adsorption ability of bare NPs on each of these metal ions in pH 2, which indicates that there is nearly no selectivity of bare NPs in acidic medium. In figure 10 (b), one can clearly see a maximum uptake of U (VI) by the polymer coated NPs in pH 7. This result shows promising potential of the polymer coated NPs in the application of U (VI) removal under neutral environment.

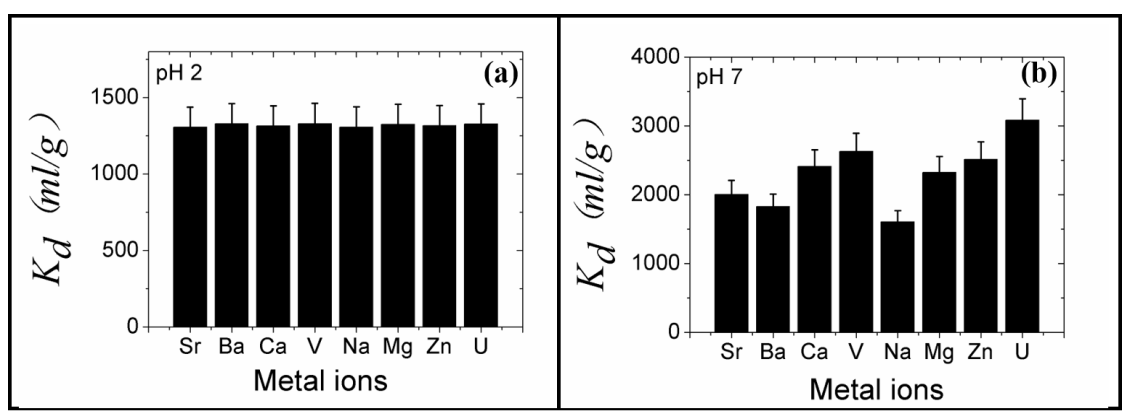

*Figure 10. Sorption distribution coefficient ($K_d$) of bare NPs on different metal ions in pH 2 (a) and of polymer coated NPs on different metal ions in pH 7 (b).*

**4, Conclusion**

In this work, bare and polymer-coated NPs were synthesized for adsorption of U (VI) from solution. In acidic pH (~ 2), the bare NPs were used to absorb the $UO_2^{2+}$ ions. The corresponding distribution coefficient $K_d$ was ~ $10^3$ ml/g. TEM and SAXS data show that the NPs aggregated after adsorption because the electrostatic balance near the particle surface charge changed in the presence of $UO_2^{2+}$. At neutral pH (~ 7), uranyl hydrolyzed and formed a plate-like hydroxide as seen in both TEM and SAXS. These plate-like hydroxide particles were 20~50 nm in diameter and ~ 5 nm thick. The PAA-coated NPs absorbed/captured these U (VI) hydroxides. TEM and SAXS evidenced the effective adsorption of PAA-coated NPs on the surface of the (VI) hydroxide. This generated bigger hybrid aggregates that can be easily extracted from aqueous medium via an external magnetic field. The corresponding distribution coefficient $K_d$ was ~ $2\times10^3$ ml/g. These results show that the bare and coated superparamagnetic NPs obtained by a simple co-precipitation procedure could be an effective absorbent for the removal of U (VI) from aqueous media.

**Authors' information**

**Corresponding authors**

yanminhao@swust.edu.cn and tuoxianguo@suse.edu.cn

**Author Contribution**

⊥S. Z and Y. L contributed equally to this work.

**Notes**

The authors declare no competing financial interest.

**Acknowledgement**

This research was supported by National Nature Science Foundation of China (Grant No. 41630646 and No. 21701157); Sichuan Province Education Department Innovation Team Foundation (16zd1104); Sichuan Province Science Foundation for Young Scientists (No. 15zs2111); Open Project of the Key Laboratory of Neutron Physics and Institute of Nuclear Physics and Chemistry, China Academy of Engineering Physics (No. 2014BB06 and No. 2017CB04); and Open Project of State Key Laboratory Cultivation Base for Nonmetal Composites and Functional Materials (No. 11zxfk26).